\documentclass[12pt]{iopart}
\usepackage{graphicx}
\usepackage{iopams}

\newcommand{\ped}[1]{\ensuremath{_{\rm #1}}}

\begin{document}
\title{A phenomenological multiband Eliashberg model for LiFeAs}
\author{G A Ummarino$^1$, Sara Galasso$^1$ and A Sanna$^2$}
\address{$^1$ Istituto di Ingegneria e Fisica dei Materiali, Dipartimento di Scienza Applicata e Tecnologia, Politecnico di Torino, Corso Duca degli Abruzzi 24, 10129
Torino, Italy}
\address{$^2$ Max-Planck-Institut f\"ur Mikrostrukturphysik, Weinberg 2, D-06120 Halle, Germany}
\ead{giovanni.ummarino@infm.polito.it}
\begin{abstract}
The phenomenology of LiFeAs superconductor can be explained in the
framework of a four-band s$\pm$-wave Eliashberg theory. We have
examined the experimental data available in literature and we have
found out that it is possible to reproduce the experimental critical
temperature, the gap values and the upper critical magnetic field within an effective model in moderate strong-coupling regime that must include both an intraband term $\lambda_{11}\sim0.9$ and an interband
spin-fluctuations ($\lambda_{tot}^{sf}\sim1.5$) coupling. The presence of a non negligible intraband coupling can be a fictitious effect of the violation of Migdal's Theorem.
\end{abstract}
\noindent{\it keywords\/}: Multiband
superconductivity, Fe-based superconductors, Eliashberg equations,
Non-phononic mechanism
\maketitle
\begin{figure}[htb]
\begin{center}
\includegraphics[keepaspectratio, width=6cm]{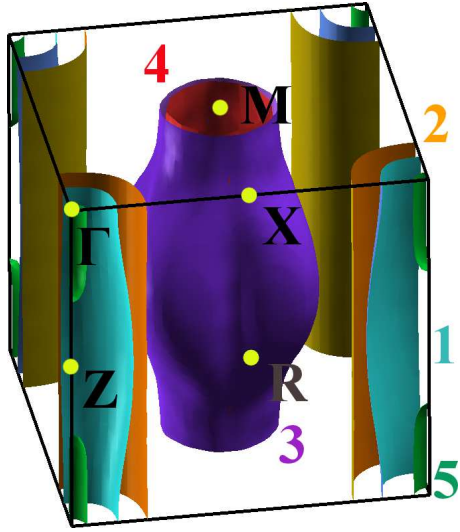}
\caption{\label{fig:FS}Fermi Surface of LiFeAs.}\end{center}
\end{figure}

%
The family of iron pnictide superconductors, discovered by the Hosono group~\cite{Hosono} in 2008,
has focused an intense research in this four years.
Many compounds with different crystal structures and physical properties have been discovered
and characterized~\cite{Stewart_rev,Mazin_rev,John_rev}, among them LiFeAs~\cite{TcTapp} has proved to be a peculiar one.\\
First of all LiFeAs, unlike almost all other iron-compounds, does
not need neither charge doping nor pressure to condense in the
superconducting state~\cite{TcTapp}, this implies that no disorder
is present. Further, it seems to be not magnetic and angle-resolved
photoemission spectroscopy (ARPES) reports poor nesting~\cite{BorisNesting}.
At a glance these characteristics could turn us
away from the idea of an unconventional pairing mechanism, however
the phonon-mediated coupling seems to be too weak~\cite{Huang} to
explain the relatively high critical temperature (T$_c$=18 K).
Moreover, recent neutron inelastic scattering measurements show the
evidence of strong antiferromagnetic fluctuations~\cite{Taylor},
reconciling LiFeAs with other Fe-based superconductors.\\
A multigap scenario is suggested by the presence of five different
sheets in the Fermi surface~\cite{Singh}: Two electron pockets are
centered near the M-point of the Brillouin zone and three hole
pockets around the $\Gamma$-point. Despite the Fermi surface shows
five different sheets, according to our electronic structure calculations the 5-th sheet can be disregarded because of its
low density of states (see Table~\ref{tab:dft} and Figure~\ref{fig:FS}) and size~\cite{computational,KS,PBE,espresso,PDdolghi}.
Consequently, as a starting point, we can model the electronic structure of LiFeAs by using a
four-band model~\cite{Ema,Popo} with two hole bands (1 and 2) and
two electron bands (3 and~4).\\
In this work we construct a theoretical model to describe the phenomenology of LiFeAs, by using a
minimal number of phenomenological parameters in combination with Density Functional Theory (DFT)
calculations and the Eliashberg theory of superconductivity. The model is then tested by simulating the temperature dependance of the critical field. \\

Recently, four different gaps have been observed by ARPES
measurements on LiFeAs~\cite{GapUmez}.
Hence, in order to describe the superconductive properties of this compound,  a four-band Eliashberg model~\cite{Ema,Popo} with \mbox{s$\pm$} symmetry~\cite{Mazin_spm} can be used. \\
The experimental gap values~\cite{GapUmez} have been used to fix the free parameters of the model
and, after this, the critical temperature and the upper critical magnetic field~\cite{Bc2Kasahara} have been calculated.
The final result is a moderate strong-coupling regime
$\lambda_{tot}\sim1.6-2.0$ where the total electron-boson coupling must
include two different contributions: a purely interband coupling
mediated by spin-fluctuations (\emph{sf}), with
$\lambda_{tot}^{sf}\sim1.5$ and a purely intraband coupling
$\lambda_{11}$, whose origin will be discussed hereafter.
\begin{table}
\caption{\label{tab:dft}Fermi Surface resolved Kohn Sham
properties~\cite{computational}: The Fermi density of states ($N(0)$)  is given in
states/spin/eV, the Fermi Velocities ($v_F$) in  $10^5$ m/sec, and
plasma frequencies ($\omega_p$) in eV. $ab$ label the in-plane and
$c$ for the out-of-plane direction of the Fermi velocities and the
diagonals of the plasma tensor~\cite{PDdolghi}.}
 \newcommand{\hs}{\hspace{0.5cm}}
 \begin{indented}
 \item[]
\begin{tabular}{@{}c|ccccc}
\br
FS  &\hs N(0)  &\hs$v_F^{\parallel ab}$\hs&$v_F^{\parallel c}$&\hs$\omega_p^{\parallel ab}$&\hs$\omega_p^{\parallel c}$\\
\mr
1   &\hs 0.556 &\hs  1.157  &\hs  0.207 &\hs  1.131  &\hs  0.202\\
2   &\hs 0.646 &\hs  1.382  &\hs  0.032 &\hs  1.455  &\hs  0.034\\
3   &\hs 0.616 &\hs  1.535  &\hs  0.865 &\hs  1.581  &\hs  0.890\\
4   &\hs 0.370 &\hs  2.014  &\hs  0.459 &\hs  1.161  &\hs  0.365\\
5   &\hs 0.039 &\hs  2.454  &\hs  1.227 &\hs  0.639  &\hs  0.319\\
\mr
TOT &\hs 2.228 &\hs  1.523  &\hs  0.529 &\hs  2.980  &\hs  1.035 \\
\br
\end{tabular}
\end{indented}
\end{table}
The isotropic values of the gaps at
$T=8$ K are reported to be $\Delta_1=5.0$~meV, $\Delta_2=2.6$~meV,
$\Delta_3=3.6$~meV, $\Delta_4=2.9$~meV. As a first approximation, since just a small anisotropy is observed, we consider only the isotropic part.\\
The Eliashberg theory~\cite{Eliashberg} generalized to multi-band
systems has already been successfully used to describe the
properties of MgB$_{2}$~\cite{Carbinicol,Carb2,Carb3} and iron compounds
\cite{Umma1,Umma2}. A four-band Eliashberg model includes eight
coupled equations for the gaps $\Delta_{i}(i\omega_{n})$ and the
renormalization functions $Z_{i}(i\omega_{n})$, where $i$ is the
band index (that ranges between $1$ and $4$) and $\omega_{n}$ are
the Matsubara frequencies.  The imaginary-axis equations are:
\begin{eqnarray}
&&\omega_{n}Z_{i}(i\omega_{n})=\omega_{n}+ \pi T\sum_{m,j}\Lambda^{Z}_{ij}(i\omega_{n},i\omega_{m})N^{Z}_{j}(i\omega_{m})+\nonumber\\
&&+\sum_{j}\big[\Gamma\ped{ij}+\Gamma^{M}\ped{ij}\big]N^{Z}_{j}(i\omega_{n})
\label{eq:EE1}
\end{eqnarray}
\begin{eqnarray}
&&Z_{i}(i\omega_{n})\Delta_{i}(i\omega_{n})=\pi
T\sum_{m,j}\big[\Lambda^{\Delta}_{ij}(i\omega_{n},i\omega_{m})-\mu^{*}_{ij}(\omega_{c})\big]\times\nonumber\\
&&\times\Theta(\omega_{c}-|\omega_{m}|)N^{\Delta}_{j}(i\omega_{m})
+\sum_{j}[\Gamma\ped{ij}+\Gamma^{M}\ped{ij}]N^{\Delta}_{j}(i\omega_{n})\phantom{aaaaaa}
 \label{eq:EE2}
\end{eqnarray}
where $\Gamma\ped{ij}$ and $\Gamma^{M}\ped{ij}$ are the non magnetic
and magnetic impurity scattering rates,
$\Lambda^{Z}_{ij}(i\omega_{n},i\omega_{m})=\Lambda^{ph}_{ij}(i\omega_{n},i\omega_{m})+\Lambda^{sf}_{ij}(i\omega_{n},i\omega_{m})$,
$\Lambda^{\Delta}_{ij}(i\omega_{n},i\omega_{m})=\Lambda^{ph}_{ij}(i\omega_{n},i\omega_{m})-\Lambda^{sf}_{ij}(i\omega_{n},i\omega_{m})$,  \emph{sf} means antiferromagnetic spin fluctuations and \emph{ph} phonons.\\
$\Theta(\omega_{c}-|\omega_{m}|)$ is the Heaviside function and
$\omega_{c}$ is a cutoff energy.
In particular,\\
$\displaystyle{\Lambda^{ph,
sf}_{ij}(i\omega_{n},i\omega_{m})=2
\int_{0}^{+\infty}d\Omega \Omega
\frac{\alpha^{2}_{ij}F^{ph,sf}(\Omega)}{(\omega_{n}-\omega_{m})^{2}+\Omega^{2}},}$
$\mu^{*}_{ij}(\omega\ped{c})$ are the elements of the $4\times 4$
Coulomb pseudopotential matrix and, finally,\\
$N^{\Delta}_{j}(i\omega_{m})=\Delta_{j}(i\omega_{m})\cdot \left[
{\sqrt{\omega^{2}_{m}+\Delta^{2}_{j}(i\omega_{m})}}\,\right]^{-1}$,\\
$N^{Z}_{j}(i\omega_{m})=\omega_{m}\cdot \left[{\sqrt{\omega^{2}_{m}+\Delta^{2}_{j}(i\omega_{m})}}\,\right]^{-1}$.\\
The electron-boson coupling constants are defined as
\begin{equation}
\lambda^{ph,sf}_{ij}=2\int_{0}^{+\infty}d\Omega\,\frac{\alpha^{2}_{ij}F^{ph,sf}(\Omega)}{\Omega}.
\end{equation}

The solution of \Eref{eq:EE1} and \Eref{eq:EE2} requires a huge
number of input parameters (32 functions and 16 constants); nevertheless,
some of these are interdependent, others may be
extracted from experiments and still others fixed by appropriate
approximations. \\
At the beginning we fixed the same conditions that have been used
for many other pnictides, as shown in Ref.~\cite{Mazin_spm}, and we
assumed that: i) the total electron-phonon coupling constant is
small and mostly intraband~\cite{Boeri}; ii) antiferromagnetic spin fluctuations mainly provide interband
coupling~\cite{Mazin_rev,Umma1}. To account for these assumptions in
the simplest way (as has already been done for other iron-compounds
with good results) we should  take:
$\lambda^{ph}_{ii}=\lambda^{ph}_{ij}=0$,
$\mu^{*}_{ii}(\omega\ped{c})=\mu^{*}_{ij}(\omega\ped{c})=0$ i.e. the
electron-phonon coupling constant and the Coulomb pseudopotential
compensate each other, in first approximation, and
$\lambda^{sf}_{ii}=0$, i.e. SF produce only interband coupling~\cite{Umma1}.
However, within these assumptions, we were not able to reproduce the
gap values of LiFeAs, and in particular the high value of
$\Delta_1$, the best results obtained are reported in the second row of Table~\ref{tab:tab2}. \\In order to solve this problem it is necessary to
introduce at least an intraband coupling in the
first band, then $\lambda_{11}\neq 0$.\\
The final matrix of the electron-boson coupling constants becomes
\begin{equation}
\lambda_{ij}= \left (
\begin{array}{cccc}
  \lambda_{11}                 &           0                   &               \lambda_{13}       &    \lambda_{14}  \\
  0                 &           0                   &              \lambda_{23}      &     \lambda_{24}   \\
  \lambda_{31}=\lambda_{13}\nu_{13}& \lambda_{32}= \lambda_{23}\nu_{23}   & 0  & 0\\
  \lambda_{41}=\lambda_{14}\nu_{14} & \lambda_{42}=\lambda_{24}\nu_{24}  & 0&  0\\
\end{array}
\right ) \label{eq:matrix}
\end{equation}
where $\nu_{ij}=N_{i}(0)/N_{j}(0)$ and $N_{i}(0)$ is the normal
density of states at the Fermi level for the $i-th$ band ($i=1,
2,3,4$).
We chose spectral functions with Lorentzian shape i.e:
\begin{equation}
\alpha_{ij}^2F_{ij}(\Omega)= C_{ij}\big\{L(\Omega+\Omega_{ij},Y_{ij})- L(\Omega-\Omega_{ij},Y_{ij})\big\}
\end{equation}
where $L(\Omega\pm\Omega_{ij},Y_{ij})=\frac{1}{(\Omega
\pm\Omega_{ij})^2+Y_{ij}^2}$ and $C_{ij}$ are normalization
constants, necessary to obtain the proper values of $\lambda_{ij}$
while $\Omega_{ij}$ and $Y_{ij}$ are the peak energies and
half-widths of the Lorentzian functions, respectively~\cite{Umma1}.
In all the calculations we set
$\Omega_{ij}=\Omega_{ij}^{sf}=\Omega_0^{sf}=8$ meV~\cite{Taylor},
and $Y_{ij}=Y_{ij}^{sf} = \Omega_{ij}^{sf}/2$~\cite{Inosov}. The
cut-off energy is $\omega_c = 18\,\Omega_0^{sf}$ and the maximum
quasiparticle energy is $\omega_{max}=21\,\Omega_0^{sf}$.
We put $\Gamma\ped{ij}$=$\Gamma^{M}\ped{ij}$=0 because the ARPES measurement are on very good single crystals of LiFeAs~\cite{GapUmez}.
Bandstructure calculations (see Table \ref{tab:dft}) provide information about
the factors $\nu_{ij}$ that enter the definition of
$\lambda_{ij}$. The obtained values are $\nu_{13}=0.9019$, $\nu_{14}=1.5010$, $\nu_{23}=1.0483$, $\nu_{24}=1.7447$.\\
After these considerations the free parameters are reduced to the five coupling
constants
$\lambda_{13},\,\lambda_{23},\,\lambda_{14},\,\lambda_{24}$ and
$\lambda_{11}$. First of all we solved the imaginary-axis Eliashberg
Equations \eref{eq:EE1} and \eref{eq:EE2} (actually we continued
them analytically on the real-axis by using the approximants Pad\`e technique)
and we fixed the free parameters in order to reproduce the gap
values at low temperature.\\

\begin{figure}[!]
\begin{center}
\includegraphics[keepaspectratio, width=8cm]{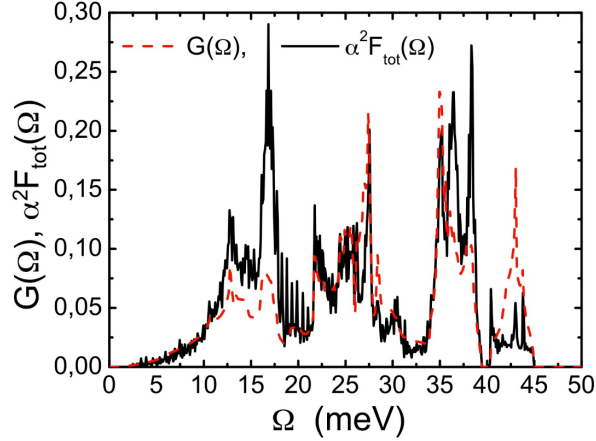}
\caption{The calculated phononic density of states
$G(\Omega)$ (red dashed line) and the calculated total
electron-phonon spectral function $\alpha^{2}F_{tot}(\Omega)$ (black
solid line).}\label{fig:phon}
\end{center}
\end{figure}

\begin{figure}[htb]
\begin{center}
\includegraphics[keepaspectratio, width=8cm]{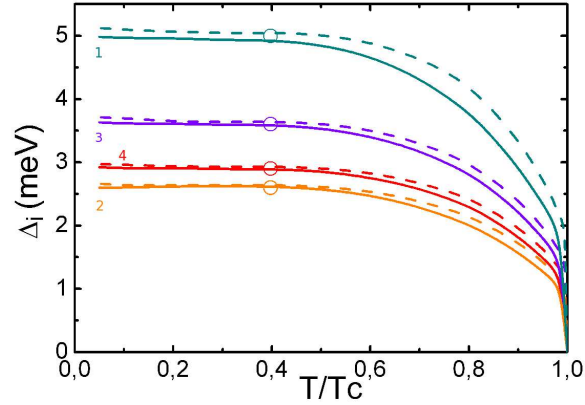}
\caption{Temperature dependence of the absolute gap
 values (\emph{lines}) and experimental data (\emph{symbols}) at 8K.
  The dark cyan solid (dashed) line represents the first gap, the orange solid (dashed) line the second one, the
  violet solid (dashed) line the third and the red solid (dashed) line the fourth, calculated with the parameters of fourth (third) row of Table~\ref{tab:tab2}.}\label{fig:gap}
\end{center}
\end{figure}

The large number of free parameters (five) may suggest that it is
possible to find different sets that produce the same results. It is
not so, as a matter of fact the predominantly interband character of
the model drastically reduce the number of possible choices. \\
At the beginning, in order to have the fewest number of free parameters, we set $\Omega_{11}$ to be the same of the
antiferromagnetic SF (even if the intraband coupling can not be mediated by SF). In this case (see third row of table~\ref{tab:tab2})
the value of $\lambda_{11}$ is very large.
We add only $\lambda_{11}$ because only the band 1 has a low Fermi energy and only in this band the Migdal's theorem breaks down.
The effect of the vertex corrections~\cite{Pietronero1,Pietronero2} can be simulated by an effective coupling that is bigger than real coupling~\cite{ema1,UmmaMigdal}.
Certainly in the bands 2, 3 and 4, the phonon couplings are very small and therefore we have not considered the possibility to have $\lambda_{22},\,\lambda_{33},\,\lambda_{44}\neq0$
and also, in a previous paper, we have demonstrated that, in these systems, the effect of a small phononic intraband coupling is negligible~\cite{Umma1}.
\\
Then we considered for $\Omega_{11}$ the typical phonon energies~\cite{Huang}. In this case (as reported in fourth and fifth row of table~\ref{tab:tab2}) the value of $\lambda_{11}$ is 0.86-0.9, while the antiferromagnetic spin fluctuations contribution correspond to a moderate strong coupling regime \mbox{($\lambda_{tot}^{sf}\sim1.5$).}\\
In fact, we have solved the Eliashberg
equations in two other cases: in the first case we used as
$\alpha_{11}^{2}F(\Omega)$ the calculated phonon density of states
$G(\Omega)$ and the second case we considered the calculated total
electron-phonon spectral function $\alpha^{2}F_{tot}(\Omega)$
both appropriately scaled. The proper choice is the second one, but this spectral function should be in principle different for each band; since the first band shows peculiar characteristics the evaluation of the electron phonon coupling could be not so reliable. Then we decided to use the phonon spectra and let the coupling to be a free parameter. These spectral functions are shown in Figure~\ref{fig:phon}.\\
At this point there are no more free parameters.
The critical temperature can be evaluated and it turns out to be
very close to experimental one,  $T^{calc}_c=18.6-20.1$~K.\\
\begin{table}
\caption{\label{tab:tab2}The first row shows the experimental data. The
second row concerns the pure interband case ($\lambda_{ii}=0$)
while the last three include an intraband term ($\lambda_{11}\neq0$): A very large value appears in the first case (the third row), a smaller one if the phonon spectral function $G(\Omega)$ (fourth row) or the electron-phonon spectral function $\alpha^{2}F(\Omega)$ (fifth row) are considered. The critical temperatures are given in K and the gap values in meV.}
\begin{indented}
\item[]\begin{tabular}{@{}c|c|ccccc|ccccc}
\br
&$\lambda_{11}$    &$\lambda_{tot}$ &$\lambda_{13}$&$\lambda_{23}$&$\lambda_{14}$&$\lambda_{24}$&$\Delta_{1}$&$\Delta_{2}$&$\Delta_{3}$&$\Delta_{4}$&$T_{c}$\\
\mr
Ex. & -& -& -& -& -& -&5.0 & 2.6  & 3.6&  2.9&18.0\\
sf &0.00   &1.80 &  1.78  &  0.66 & 0.45  & 0.52  & 3.7 & 2.6  & 3.6&  2.9&  15.9\\
sf,? &2.10    & 2.00 & 1.15 & 0.80 &  0.45  & 0.30  &5.0 & 2.6  & 3.6&  2.9&  18.6\\
sf, ph (1)	&0.86    & 1.62 & 1.06 & 0.79 &  0.42  & 0.30  &5.1 & 2.6  & 3.7&  2.9&  20.0\\
sf, ph  (2)& 0.90    & 1.63 & 1.15 & 0.80 &  0.45  & 0.30  &5.0 & 2.6  & 3.6&  2.9&  20.1\\
\br
\end{tabular}
\end{indented}
\end{table}
By solving real-axis Eliashberg equations we obtained the
temperature dependence of the gaps (see Figure~\ref{fig:gap}) for the
parameter sets reported in the third and fourth row of Table~\ref{tab:tab2}. \\
\begin{figure}[htb]
\begin{center}
\includegraphics[keepaspectratio, width=8cm]{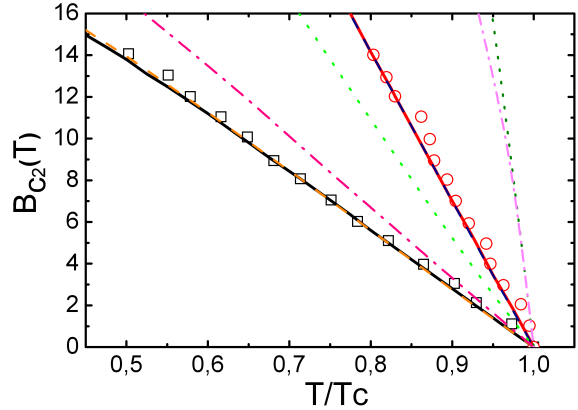}
\caption{Experimental temperature dependence of the
upper critical field    (symbols), and the relevant fitting curves
(lines) obtained by solving the Eliashberg equations in magnetic
field. Red circles and solid red (dashed dark blue) line for $H_{ \parallel c}$, black square
symbols and solid black (dashed orange) line for $H_{ \parallel ab}$ calculated with the parameters of fourth (third) row of Table~\ref{tab:tab2}.
The dotted olive ($H_{ \parallel c}$) and navy ($H_{ \parallel ab}$) and the dashed-dotted magenta ($H_{ \parallel c}$) and pink ($H_{ \parallel ab}$) lines are,respectively, the fourth and first case of Table~\ref{tab:tab2} but where $v_{F1}$ isn't a free parameter but it is taken by Table~\ref{tab:dft}}\label{fig:Bc2}
\end{center}
\end{figure}
The multiband Eliashberg model developed above can also be used to
explain the experimental data of temperature dependence of the upper critical magnetic field~\cite{Bc2Kasahara}.
For the sake of completeness, we give here the linearized gap
equations in the presence of magnetic field, for a superconductor
in the clean limit. In the following, $v_{Fj}$ is the Fermi velocity
of the $j-th$ band, and $H_{c2}$ is the upper critical field:
\begin{eqnarray}
\omega_{n}Z_{i}(i\omega_{n})&=&\omega_{n}+\pi
T\sum_{m,j}\Lambda^{Z}_{ij}(i\omega_{n}-i\omega_{m})\mathrm{sign}(\omega_{m})\\
Z_{i}(i\omega_{n})\Delta_{i}(i\omega_{n})&=&\pi
T\sum_{m,j}\left[\Lambda^{\Delta}_{ij}(i\omega_{n}-i\omega_{m})-\mu^{*}_{ij}(\omega_{c})\right]\cdot \nonumber\\
& &
\cdot\theta(|\omega_{c}|-\omega_{m})\chi_{j}(i\omega_{m})Z_{j}(i\omega_{m})\Delta_{j}(i\omega_{m})
\end{eqnarray}
\begin{eqnarray}
\chi_{j}(i\omega_{m})&=&\frac{2}{\sqrt{\beta_{j}}}\int^{+\infty}_{0}dq\exp(-q^{2})\cdot\nonumber\\
& &\cdot
\tan^{-1}\left[\frac{q\sqrt{\beta_{j}}}{|\omega_{m}Z_{j}(i\omega_{m})|+i\mu_{B}H_{c2}\mathrm{sign}(\omega_{m})}\right].\nonumber
\end{eqnarray}
\noindent Here $\beta_{j}=\pi H_{c2} v_{Fj}^{2}/(2\Phi_{0})$ and
$\Phi_{0}$ is the unit of magnetic flux. In these equations the four
bare Fermi velocities $v_{Fj}$~\cite{Suderow} are the input
parameters. As the first band shows peculiar characteristics even in the calculation of the Fermi velocity can be present some corrections, then we decided to let the first Fermi velocities to be free parameters and we choose them to find the
best fit of the experimental data~\cite{Bc2Kasahara} while the other
values have been fixed to the values reported in Table~\ref{tab:dft}.
Then $v_{F1}$, in each case,  is the only free
parameter. The obtained values are: $v_{F1}^{\parallel c}=
2.28\cdot10^5$ m/s and $v_{F1}^{\parallel ab}= 1.74\cdot10^5$ m/s, in the phonon case and
$v_{F1}^{\parallel c}=2.79\cdot10^5$ m/s and $v_{F1}^{\parallel ab}= 2.14\cdot10^5$ m/s, if the spin fluctuation spectral function is considered. Figure~\ref{fig:Bc2} shows the experimental data and the
best theoretical curves (solid and dashed lines) obtained by solving the Eliashberg equations
within the model discussed above. As can be seen, the results obtained in the two considered cases are almost indistinguishable and in very good agreement with the experimental data.
In the Figure~\ref{fig:Bc2} we shows also the curves calculated with $v_{F1}$ taken by Table~\ref{tab:dft} when $\lambda_{11}\neq0$ (fourth case in Table~\ref{tab:tab2}, dotted line olive and navy) and when $\lambda_{11}=0$ (first case in Table~\ref{tab:tab2}, dashed-dotted line magenta and pink). In both situations there is no agreement with the experimental data. The curves calculated in absence of the term $\lambda_{11}$ don't agree with the experimental data so we deduce that the higher value of $v_{F1}^{\parallel ab,c}$ isn't produced by the presence of an intraband term ($\lambda_{11}\neq0$) but, probably, by the peculiar characteristics of band 1. However one must consider the fact that the Eliashberg equations are derived by assuming Migdal's theorem. In presence of an anomalous band dispersion as for band 1, the theory may partially break down. Allowing $v_{F1}^{\parallel ab,c}$ as a free parameter implicitly implies that we are "phenomenologically" going beyond the first order contributions (i.e. now we cannot neglect the effects of the vertex corrections in the band 1).
The break down of the Migdal's theorem leads to use effective values of $\lambda_{11}$ and $v_{F1}$ different from real value because the framework of the theory is partially inadequate.

To summarize we have constructed a phenomenological model of superconductivity for LiFeAs able to describe its critical temperature, the multigap structure measured by Umezawa and coworkers in Ref.~\cite{GapUmez} and other experimental observations.
However this process was not straightforward, to be able to conjugate a spin fluctuation dominated pairing with the experimental gap structure we have been forced to introduce an intraband coupling that acts only on the first band.
The presence of a phononic, purely intraband term seems to indicate an intrinsic incompatibility between the structure of the superconducting gaps as measured by Umezawa and coworkers in Ref.~\cite{GapUmez} and a purely spin-fluctuation mediated pairing.
A possible explanation may be linked to the very low Fermi energy of the band for which vertex corrections~\cite{Pietronero1,Pietronero2} to the usual Migdal-Eliashberg theory may be relevant, and are expected to increase the stenght of the phononic pairing~\cite{ema1,UmmaMigdal}.\\
In conclusion, our calculations show that LiFeAs presents peculiar features with respect to other iron compounds and it cannot be explained within the framework of a pure interband spinfluctuation mediated superconductivity.




\end{document}